\documentclass[cameraready]{Interspeech}
\usepackage{tabularx}
\usepackage{threeparttable}

\title{Multi-View Based Audio Visual Target Speaker Extraction}

\author[affiliation={1,2}, orcid=0009-0004-4948-2028 ]{Peijun}{Yang}
\author[affiliation={3}, orcid=0009-0002-8157-7970]{Zhan}{Jin}
\author[affiliation={2,3}, orcid=0000-0001-9344-7415 ,correspondingauthor]{Juan}{Liu}
\author[affiliation={2,4}, orcid=0000-0002-6406-1983 ,correspondingauthor]{Ming}{Li}

\address{
    $^1$Key Laboratory of Aerospace Information Security and Trusted Computing, \\
    Ministry of Education, School of Cyber Science and Engineering, Wuhan University, Wuhan, China \\
    $^2$School of Artificial Intelligence, Wuhan University, Wuhan, China\\
    $^3$School of Computer Science, Wuhan University, Wuhan, China \\
    $^4$School of Artificial Intelligence, The Chinese University of Hong Kong, Shenzhen, China
}

\email{liujuan@whu.edu.cn,ming.li.cuhksz@gmail.com}


\usepackage{comment}

\begin{document}

\maketitle

\begin{abstract}
Audio-Visual Target Speaker Extraction (AVTSE) aims to separate a target speaker's voice from a mixed audio signal using the corresponding visual cues. While most existing AVTSE methods rely exclusively on frontal-view videos,
this limitation restricts their robustness in real-world scenarios where non-frontal views are prevalent.
Such visual perspectives often contain complementary articulatory information that could enhance speech extraction. In this work, we propose Multi-View Tensor Fusion (MVTF), a novel framework that transforms multi-view learning into single-view performance gains.
During the training stage, we leverage synchronized multi-perspective lip videos to learn cross-view correlations through MVTF, where pairwise outer products explicitly model multiplicative interactions between different views of input lip embeddings.
At the inference stage, the system supports both single-view and multi-view inputs. Experimental results show that in the single-view inputs, 
our framework leverages multi-view knowledge to achieve significant performance gains, while in the multi-view mode, it further improves overall performance and enhances the robustness.
Our demo, code and data are available at \footnote{https://anonymous.4open.science/w/MVTF-Gridnet-209C/}
\end{abstract}


\section{Introduction}
Target Speaker Extraction (TSE) separates a specific speaker's voice from overlapping speech or background noise, making it essential for hearing aids and speech recognition systems.
Traditional TSE approaches typically rely on auxiliary reference cues to direct attention toward the target speaker. 
Audio-only methods utilize the target speaker's characteristics from a pre-recorded speech \cite{xu2020spex,hao2024x,ge2020spex+},
while visually-guided approaches usually leverage the synchronized visual information \cite{pan2021muse,afouras2018conversation,pan2023scenario} such as lip movements or auxiliary face, hence enrollment is not needed.
Recently, visual cue based methods become more popular, as they do not rely on pre-registered information and remain unaffected by highly noisy acoustic interference.
Most AVTSE systems follow an encoder-fusion-separator-decoder pipeline. Audio and visual encoders extract embeddings; the latter are upsampled and fused—typically via concatenation \cite{wu2019time,kalkhorani2025av,pan2022usev} or advanced strategies \cite{li2024audio,li2023iianet}—before a separator network estimates a mask or target features, which is decoded into the target waveform. Common separators include GridNet \cite{wang2023tf}, CrossNet \cite{kalkhorani2024tf}, and DPRNN \cite{luo2020dual}.

However, most existing visual-guided AVTSE systems predominantly assume the availability of frontal facial views. 
This assumption is reflected in widely-used datasets such as LRS3 \cite{afouras2018lrs3} and VoxCeleb2 \cite{chung2018voxceleb2},
where speakers are typically captured from a frontal perspective.
Since these datasets lack annotations for head pose variations, it remains unclear whether training and validation on stable frontal views is more effective than leveraging multiple distinct perspectives.
Moreover, when synchronized multi-view video of a speaker is available, it raises the question of how to effectively exploit such synchronized information across different angles to further enhance model performance. 
Also, the reliance on frontal views significantly limits the practical applicability of AVTSE systems in real-world scenarios, 
where speakers naturally rotate their heads or cameras capture non-frontal angles—factors that have been shown to degrade performance \cite{jordan1997seeing}.

Previous work generates pose-invariant information \cite{liu2023piave} or robust face frontalization\cite{kang2022impact,kang2021robust} for reducing the impact of head movements.
Attempting to correct non-frontal views rather than learning to leverage it may discard the original visual information, which results in suboptimal performance when frontalization fails.
Some works in other related fields attempt to identify and process each specific pose separately \cite{hesse2012multi} or use multiple cameras' information to enhance performance \cite{petridis2017end}.
The former is limited by the number of poses,
while the latter requires a fixed multi-camera setup in both training and testing.
To address these limitations, we propose a novel multi-view learning framework.
Instead of treating pose variation as a problem to be corrected via frontalization, 
we embrace it as a source of complementary information.
The key insight is that synchronized multi-view videos inherently contain complementary articulatory cues—learning such cross-view correlations directly enhances each individual view's representation, 
making it more robust to pose variations.
Our framework learns a view-invariant and robust visual representation by explicitly modeling the multiplicative interactions between different viewing angles during training, 
rather than relying on the availability of multiple cameras at test time.
Critically, this multi-view training strengthens the representation for each individual view.
This is because the model learns to extract shared articulatory information across perspectives, enriching the feature space for each view rather than simply averaging or selecting among them.
The model leverages multi-view data during training yet accepts single-view inputs at test time, making it practical for real-world single-camera scenarios.
When tested on a video of a speaker freely turning their head, it applies learned cross-view knowledge to compensate for continuous pose changes—a capability frontalization methods cannot achieve.

\begin{figure*}[htbp]
  \centering
  
  \includegraphics[scale=0.2]{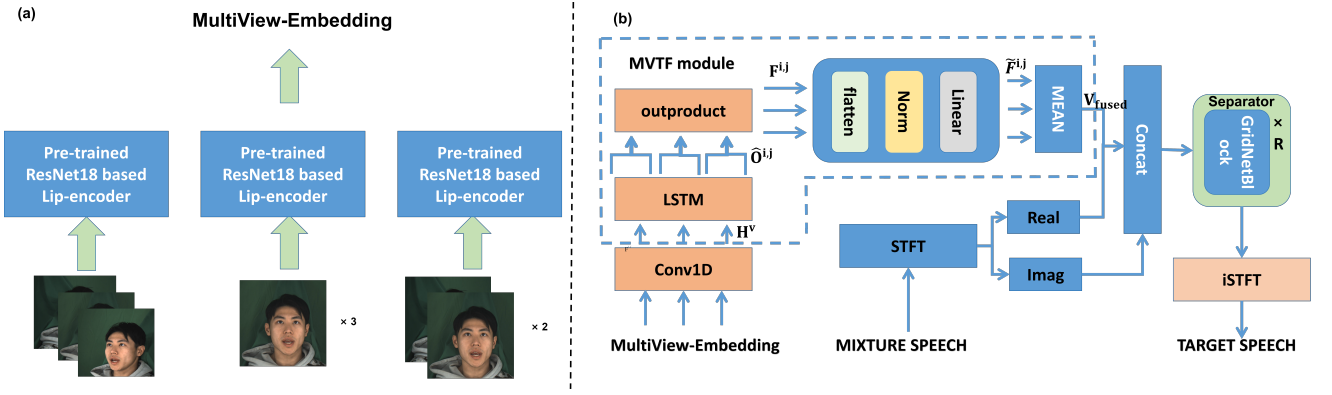}
  \caption{Our proposed MVTF method that extracts the target speakers' speech using Multi-View lip embeddings.
  (a)Three views, one view (copied three times), and two views (one view copied two times) visible situations. (b)overall structure of MVTF-GridNet.}
  \label{figure1}
\end{figure*}
\section{Methods}

\subsection{Overview}
As shown in Fig\ref{figure1}, we propose a robust AVTSE system that utilizes multi-view visual data in training to achieve superior performance during the inference even with a single (typically frontal) view testing video.
Based on the TF-GridNet backbone \cite{wang2023tf}, our model introduces a Multi-View Tensor Fusion module.
The mixed audio is encoded by the Short-Time Fourier Transform (STFT), while lip encoders \cite{martinez2020lipreading} process each video view. The MVTF module employs Long Short-Term Memory networks (LSTMs) \cite{hochreiter1997long} and a tensor outer product to fuse multi-view embeddings into a view-invariant representation.
This is combined with audio features for the GridBlock to reconstruct the target speaker's spectrogram, which is then converted to a waveform  through iSTFT.
\subsection{Audio Processing Backbone}

We adopt TF-GridNet \cite{wang2023tf} as our audio separation backbone due to its strong performance in spectral mapping.
The mixture audio signal $\mathbf{X} \in \mathbb{R}^{B \times N \times M}$ (where $B$ is the batch size, $N$ is the number of waveform samples, and $M$ is the number of microphones which defaults to 1) is first normalized by its standard deviation across channels and time to ensure stable training: $\mathbf{\tilde{X}} = \mathbf{X} / \sigma(\mathbf{X})$.
A Short-Time Fourier Transform (STFT) encoder then projects the normalized waveform into a complex-valued spectrogram:
$\mathbf{C}_{mix} = \text{STFT}(\mathbf{X}) \in \mathbb{C} ^{B\times M \times  T_a\times F}$
where $T_a$ and $F$ represents the number of time frames and frequency bins, respectively. 
The real and imaginary components of $\mathbf{C}_{mix}$ are concatenated to form the input feature map $\mathbf{A} \in \mathbb{R}^{B \times 2*M \times T_a \times F}$ for the subsequent network.
\begin{table*}[t]
  \centering
  \begin{threeparttable}
    \caption{Performance of MVTF-GridNet with single-view input and other fusion strategies on the MEAD dataset. ``Training'' denotes the lip embedding strategy used during training.
    For MVTF-GridNet, ``Random'' selects 3 distinct views out of 7 per batch; ``Repeat'' selects 1 view and repeats it three times; ``Front'' uses only frontal view repeated.
    For GridNet, ``Random'' selects 1 out of 7 views per batch.
    Simple Addition and Attention Fusion are trained with the ``Random'' strategy.}
    \label{table1}
    \scriptsize
    \setlength{\tabcolsep}{4pt}
    \begin{tabularx}{\textwidth}{r l X c *{7}{c} c}
      \toprule
      & & & & \multicolumn{7}{c}{Testing View} & \\
      \cmidrule(lr){5-11}
      ID. & Model & Training & Metric & Front & Top & Down & Left30 & Left60 & Right30 & Right60 & AVG(7) \\
      \midrule
      - & Mixture & -- & SI-SDR & \multicolumn{8}{c}{-0.191} \\
      \midrule
      1 & MVTF-GridNet & Random (random 3 out of 7 views) & SI-SDR & \textbf{15.836} & \textbf{15.196} & \textbf{15.784} & \textbf{15.839} & \textbf{15.792} & \textbf{15.822} & \textbf{15.758} & \textbf{15.718} \\
      2 & MVTF-GridNet & Repeat (random 1 out of 7 views)*3 & SI-SDR & 15.133 & 14.878 & 15.202 & 15.108 & 15.138 & 15.146 & 15.109 & 15.102 \\
      3 & MVTF-GridNet & Front *3 & SI-SDR & 14.930 & 10.230 & 14.841 & 14.778 & 14.456 & 14.761 & 14.723 & 14.102 \\
      4 & GridNet      & Random (random 1 out of 7 views) & SI-SDR & 15.259 & 14.765 & 15.048 & 15.055 & 15.107 & 15.214 & 15.175 & 15.089 \\
      5 & GridNet      & Front & SI-SDR & 13.290 & 7.731 & 11.159 & 13.622 & 13.321 & 13.810 & 13.914 & 12.406 \\
      - & MVTF-GridNet & Random (random 3 out of 7 views) & SDR\tnote{†} & 10.784 & 10.646 & 10.865 & 10.873 & 10.971 & 10.801 & 10.740 & 10.811 \\
      - & PIAVE\cite{liu2023piave} & -- & SDR\tnote{†} & 11.773 & 8.923 & 8.514 & 8.583 & 6.118 & 8.387 & 4.935 & 8.176 \\
      \midrule
      6 & Projected Addition & Random & SI-SDR & 14.733 & 14.191 & 14.764 & 14.629 & 14.588 & 14.711 & 14.523 & 14.591 \\
      7 & Attention Fusion   & Random & SI-SDR & 13.938 & 13.931 & 13.937 & 13.940 & 13.937 & 13.938 & 13.936 & 13.938 \\
      \bottomrule
    \end{tabularx}
    \begin{tablenotes}[para,flushleft]
      \item[†]Although we adopt the same dataset, data splits, mixture generation, and SNR range as PIAVE~\cite{liu2023piave}, utterance-level splits may differ. Furthermore, PIAVE is pre‑trained on LRS3 and fine‑tuned on MEAD, while our models are trained from scratch. Thus, results are indicative rather than strictly comparable.
    \end{tablenotes}
  \end{threeparttable}
\end{table*}
\subsection{Multi-View Feature Extraction}
The multi-view video input $\mathbf{V}\in \mathbb{R}^{B\times V\times T_v}$ (where $B$ is the same batch size as the audio signal, $V$ is the number of views and $T_v$ is the number of video frames) is processed separately.
For each view $V_i\in \mathbb{R}^{B\times T_v}$, we employ a pre-trained lip-reading network \cite{martinez2020lipreading} to extract spatio-temporal features from the lip ROI sequence $\mathbf{V}^v$.
Prior to feature extraction, an affine transformation is applied to align each ROI to a mean face template. 
The resulted embedding $\mathbf{L}^v \in \mathbb{R}^{B \times T_v \times D}$ ($D=512$ here) provides a high-dimensional representation of the lip motion.

A critical challenge is the temporal misalignment between audio frames ($T_a$) and video frames ($T_v$). To address this, we apply linear interpolation to upsample the visual sequence to match the audio temporal resolution:
  $\tilde{\mathbf{L}} ^{v} = \text{Interpolate}(\mathbf{L}^{v},size=T_a) \in \mathbb{R}^{B\times T_a \times D}$.
This ensures that each audio frame has a corresponding visual feature vector.
Finally, we project the visual features into a common subspace shared with audio features using a 1D convolution layer:
  $\mathbf{H}^v = \text{Conv1D}(\tilde{\mathbf{L}}^v) \in \mathbb{R}^{B \times T_a \times F}$.
\subsection{Multi-View Tensor Fusion Module}
This module is designed to integrate information from multiple views even when some are missing during inference.
In the event of missing perspectives during inference, our method replicates the available views to compensate for the absent ones.
The features $\mathbf{H}^v$ are fed into a shared, single-layer LSTM to capture temporal dependencies within each view's sequence:
\begin{equation} \label{eq4}
  \mathbf{O} ^v = \text{LSTM}(\mathbf{H}^v)
\end{equation}
We use the output sequence $\mathbf{O}^v \in \mathbb{R}^{B \times T_a \times F}$ for further fusion.
We consider the lip embedding from different perspectives as multi-view but also shares a lot of similarity.
Directly integrating multi-view inputs may introduce noise that degrades training. 
Inspired by the idea of the tensor fusion network \cite{zadeh2017tensor} and the idea of fusing imperfect modality \cite{liang2019learning}, we design the Multi-View Tensor Fusion Module to fuse the lip embedding of different views to suppress noise and enhance the performance of single view properly.
To model rich multiplicative interactions between views—which additive concatenation is incapable of capturing—we compute pairwise outer products of the LSTM-processed features.
First, we augment each feature vector with a constant 1 to incorporate bias terms. After constant 1 is appended, unimodal and bimodal interactions are all captured as described in \cite{zadeh2017tensor}
\begin{equation} \label{eq5}
   \mathbf{\hat{O}} =[\mathbf{O},\mathbf{1}] \in \mathbb{R}^{B\times T_a \times (F+1)}
  \end{equation}
The fused representation for a pair of views $(i, j)$ is computed as:
\begin{equation} \label{eq6} \mathbf{F} ^{i,j} = \mathbf{\hat{O_{i}}} \otimes \mathbf{\hat{O_{j}}}
\end{equation}
where $\otimes$ denotes the outer product applied along the feature dimension for each aligned time step independently, resulting in $\mathbf{F}^{i,j} \in \mathbb{R}^{B \times T_a \times (F+1) \times (F+1)}$, which captures multiplicative interactions between the two view embeddings.

The pairwise tensors $\mathbf{F}^{i,j}$ are flattened and projected back to the original dimension $F$ using a LayerNorm and a linear layer to maintain tractability.
\begin{equation} \label{eq7} 
  \tilde{\mathbf{F}}_{i,j} = Linear(LayerNorm(Flatten(\mathbf{F}_{i,j})))
\end{equation}
The final fused visual context $\mathbf{V}_{\text{fused}}$ is obtained by averaging the contributions from all available view pairs:
\begin{equation} \label{eq8}  \mathbf{V}_{fused} = \frac{1}{\left | \cal{P}\right |} \sum_{(i,j) \in {\cal{P}}} \tilde{\mathbf{F}}^{i,j} \in \mathbb{R}^{B\times T_a\times F}
\end{equation}
where $\mathcal{P}$ is the set of all pairs of available views (The interaction between one view and itself is allowed so that we can use multi-view information to train and use only one view to infer). In our experiments, $\left | \cal{P}\right |$ was set to 3 which means we have 3 available views and pairwise interactions produce 3 results, This tensor is unsqueezed to $\mathbb{R}^{B \times 1 \times T_a \times F}$ to be combined with the audio features.
Due to the rotation symmetry of the MVTF module, the order of inputs has no effect on the results. The advantage of the symmetry is that the camera positions in the test environment do not need to be the same as those in training and can be freely matched.

\section{Experimental Setup}
\subsection{Dataset}
\label{dataset}
We conduct our experiments on the MEAD \cite{wang2020mead} dataset, a multi-view audio-visual emotional dataset. To isolate the impact of view variation from emotional variance, we exclusively use videos with neutral emotion for all speakers.
The dataset provides synchronized audio and video from 7 different camera angles (front, top, down, left 30°, right 30°, left 60°, right 60°).
For each mixture, the target speaker and an interfering speaker are randomly selected from different 
speakers in the dataset. All mixtures are a mix of two speakers.

A total of 10,000, 1,000, and 1,000 audio mixtures were generated for training, validation, and testing, respectively, ensuring that there are no overlap between the training, validation and testing partitions at either the utterance or speaker level. 
Their clean utterances are mixed at a random Signal-to-Noise Ratio (SNR) between -10 dB and 10 dB. All audio signals are resampled to 16 kHz.
The video frames are processed at 25 frames-per-second(FPS).

In real-life scenarios, the head posture is constantly changing, so it is difficult to obtain stable single-view information. 
In order to verify the robustness of the model under real-life head posture changes, 
we construct a test set from the same 1,000 test mixtures by injecting feature segments from alternate viewpoints (e.g., 30° left) into frontal view sequences to test model's robustness, while keeping the corresponding audio mixtures unchanged.
The inserted segments varied in length from 20\% to 40\% of the total sequence and were placed randomly between the 30\% and 80\% positions of the timeline.

\subsection{Implementation Details}
The proposed model was implemented using the PyTorch framework. 
The Adam \cite{diederik2014adam} optimizer was employed for training with an initial learning rate of 1e-3.
A reduction-based learning rate scheduler was utilized, which halved the learning rate if no performance improvement on the validation set was observed for three consecutive epochs. 
Training was terminated early if no improvement occurred for ten consecutive epochs. 
To ensure training stability, gradient clipping was applied, constraining the L2-norm of the gradients to a maximum value of 1.
The models were trained for a maximum of 100 epochs and optimized using a Scale-Invariant Signal-to-Distortion ratio (SI-SDR) \cite{le2019sdr} loss.

\subsection{Baseline and Ablation Studies}
To comprehensively evaluate the proposed MVTF module, 
we construct a hierarchical set of baselines and ablations.
All models share the same TF-GridNet backbone, optimization settings, and evaluation protocol. 
The comparisons span two dimensions: the presence of the MVTF module and the input strategy during training, as detailed below.

We first establish two MVTF-free baselines: GridNet (Front), trained solely on frontal views, and GridNet (Random), trained on random single views per batch to isolate the effect of viewpoint diversity.
Next, to assess the contribution of the MVTF module itself, we train MVTF-GridNet under three input strategies: 
Random (three distinct and random views out of seven views per batch), 
Repeat (one random view out of seven views per batch and repeat it three times), 
and Front \* 3 (only frontal view repeated). 
Finally, to validate the effectiveness and necessity of the proposed MVTF module, we compare MVTF against two alternative fusion strategies under the same multi‑view training setup:
Projected Addition (element\-wise summation after linear projection, capturing only additive interactions) and 
Attention Fusion (cross\-attention with random QKV assignment, offering flexibility but potential instability).

All models are evaluated on the MEAD test set with single\-view inputs.
For MVTF-GridNet, the single\-view inputs are repeated 3 times to meet the number of input channels.
Results are summarized in Table \ref{table1}


\section{Results and Discussion}
The performance of all systems was quantitatively evaluated using Scale-Invariant Signal-to-Distortion Ratio (SI-SDR) \cite{le2019sdr}, Perceptual Evaluation of Speech Quality (PESQ) \cite{pesq} and the Short-Term Objective Intelligibility (STOI) \cite{stoi}.

\subsection{Multi-View Training Enhancement for Single-View Testing}

As shown in Table\ref{table1}, 
MVTF-GridNet trained on random multi-view data achieves the best average SI-SDR of 15.718 dB, outperforming its frontal-only counterpart by 1.616 dB.
The gain is especially significant under challenging views such as the top view, demonstrating the method's effectiveness in handling non-frontal perspectives. 
The performance gap between Random (ID 1) and Repeat (ID 2) or Front (ID 3) strategies imply that exposure to diverse viewpoints during training is critical for generalization.
This is also confirmed by the fact that all models trained with random viewpoints perform better than those trained with only frontal faces. 
Interestingly, GridNet trained only on frontal views (ID 5) performs worse on frontal than on some profile views (Right30/60),may be attributed to the pre-trained visual encoder's inherent robustness to certain angles.
However, MVTF-GridNet exhibits superior robustness. 
\begin{table}[htbp]
  \centering
  \caption{Robustness to mixed frontal/non-frontal views. 
  Models (ID 1, 4, 5) correspond to MVTF-GridNet (Random), GridNet (Random), and GridNet (Front) from Table\ref{table1}, respectively.}
  \label{table2}
  \footnotesize
  \begin{tabularx}{\columnwidth}{lXXX}
    \toprule
    Model & SI-SDR & PESQ & STOI \\
    \midrule
    MVTF-GridNet (ID 1) & 15.834 & 2.887 & 0.915 \\
    GridNet-Random (ID 4) & 15.179 & 2.778 & 0.902 \\
    GridNet-Front (ID 5) & 10.425 & 2.470 & 0.851 \\
    \bottomrule
  \end{tabularx}
\end{table}

When evaluated on mixed-view sequences simulating head rotations (Table \ref{table2}), MVTF-GridNet (ID 1) maintains stable performance, while GridNet (ID 4,5) shows degradation—especially ID 5. This indicates that MVTF effectively learns cross-view correlations, improving stability over naive multi-view or single-view training.

\subsection{Performance for Multi-View Conditions during Inference}

Table \ref{table3} demonstrates the inference performance of typical multi-view combinations in real-world settings, underscoring the robustness and adaptability of our method. 
All configurations consistently achieve high performance, with SI-SDR around 15.85 dB, PESQ near 2.90, and STOI consistently above 0.915. 
\begin{table}[htbp]
  \centering
  \caption{Performance of typical multi-view input combinations. The test model is ID 1 of table \ref{table1}. The test set is the same 1,000 test mixtures by using the corresponding view combination .}
  \label{table3}
  \footnotesize
  \begin{tabular}{lccc}
    \toprule
    Combination & SI-SDR & PESQ & STOI \\
    \midrule
    Front\&Left30\&Right30 & 15.857 & 2.898 & 0.915 \\
    Front\&Left60\&Right60 & 15.854 & 2.898 & 0.916 \\
    Front\&Right30\&Right30 & 15.834 & 2.892 & 0.915 \\
    Front\&Right60\&Right60 & 15.835 & 2.894 & 0.915 \\
    Front\&Left30\&Right60 & 15.851 & 2.899 & 0.916 \\
    \bottomrule
  \end{tabular}
\end{table}
This consistency indicates that the model performs reliably across various angle arrangements without dependence on a specific configuration. 
Such flexibility is particularly beneficial in practical scenarios where multiple cameras are fixed while the target speaker moves.
\subsection{Comparison of Fusion Strategies and Model Complexity}
\begin{table}[htbp]
  \centering
  \caption{Model parameter count and computational cost (4-second mixture input).}
  \label{tab:complexity}
  \footnotesize
  \begin{tabularx}{\columnwidth}{lXXX}
    \toprule
    Model & Params (M) & FLOPs (G) & SI-SDR (dB) \\
    \midrule
    GridNet (Single-View) & 7.235 & 470.750 & 15.089 \\
    MVTF-GridNet (Ours)   & 7.561 & 471.824 & 15.718 \\
    Projected Addition     & 7.274 & 470.934 & 14.591 \\
    Attention Fusion       & 7.269 & 470.921 & 13.938 \\
    \bottomrule
  \end{tabularx}
\end{table}
We compare MVTF with Projected Addition and Attention Fusion under identical settings (ID 6–7). MVTF achieves the highest SI-SDR among the three.
Both alternatives underperform relative to the random single-view GridNet baseline: Projected Addition fails to capture nonlinear interactions, while Attention Fusion suffers from unstable cross-attention due to random QKV assignment.
This highlights the challenge of fusing multi-view information without introducing noise.
In contrast, MVTF explicitly models multiplicative interactions via outer products, enabling more effective fusion.
MVTF-GridNet adds only marginal parameters and FLOPs over the single-view baseline (Table \ref{tab:complexity}), 
yet yields substantial gains—striking an excellent trade-off between complexity and performance.

\subsection{Comparison with previous works on the MEAD dataset}
We further compare with PIAVE~\cite{liu2023piave}, a recent AVTSE model that generates pose-invariant frontal faces. As shown in Table~\ref{table1}, our MVTF-GridNet achieves an average SDR of 10.81 dB across seven views, outperforming PIAVE's 8.18 dB. 
This gap underscores the effectiveness and stability of our explicit multi-view fusion mechanism combined with the TF-GridNet\cite{wang2023tf} backbone for target speaker extraction under varying viewpoints.



\section{Conclusion}
This paper presents MVTF, a novel Multi-View Tensor Fusion framework for AVTSE. 
By explicitly modeling multiplicative interactions across different viewpoints, 
our method captures complementary articulatory information from multi-view lip videos during training, 
enabling robust performance with both multi-view and single-view inputs at test time.
Experimental results on the MEAD dataset demonstrate that MVTF outperforms frontal-only and naive multi-view baselines, particularly on challenging non-frontal views. 
The framework's flexible input configuration makes it well-suited for real-world scenarios with varying camera angles.
\section{Generative AI Use Disclosure}
Generative AI tools were used for limited language editing purpose, including improving clarity and correcting grammatical issues.
No substantive content, analysis, or conclusions were generated by AI. The authors remain fully responsible for the content of this manuscript.
\bibliographystyle{IEEEtran}
\bibliography{mybib}

\end{document}